# A numerical approach for 3D manufacturing tolerances synthesis

*Frédéric VIGNAT, François VILLENEUVE*
*G-SCOP Laboratory, Grenoble University, 46 avenue Félix Viallet,*
*38031 Grenoble Cedex 1*
*frederic.vignat@inpg.fr, francois.villeneuve@hmg.inpg.fr*

**Abstract:** Making a product conform to the functional requirements indicated by the customer suppose to be able to manage the manufacturing process chosen to realise the parts. A simulation step is generally performed to verify that the expected generated deviations fit with these requirements. It is then necessary to assess the actual deviations of the process in progress. This is usually done by the verification of the conformity of the workpiece to manufacturing tolerances at the end of each set-up. It is thus necessary to determine these manufacturing tolerances. This step is called "manufacturing tolerance synthesis". In this paper, a numerical method is proposed to perform 3D manufacturing tolerances synthesis. This method uses the result of the numerical analysis of tolerances to determine influent mall displacement of surfaces. These displacements are described by small displacements torsors. An algorithm is then proposed to determine suitable ISO manufacturing tolerances.

1. INTRODUCTION

In this paper, a method for qualitative and quantitative tolerance synthesis is proposed, in the field of manufacturing tolerancing. The qualitative synthesis is defined by convert functional (or design) tolerances into manufacturing tolerances, i.e., the tolerances of intermediate working dimensions in part manufacturing processes, including the identification of the defect parameters to be limited for each manufacturing set-up and the proposal for corresponding ISO tolerances. The quantitative synthesis is the determination of a set of values for manufacturing tolerances making it possible to guarantee the respect of the studied functional tolerance. The second stage of the quantitative synthesis consists in a choice of values making it possible to minimize the cost of the selected manufacturing process. This second stage of optimization will not be treated in this paper.

In the proposed method, the tolerance synthesis comes after the analysis step. The process has been verified as capable to produce part fitting with the functional tolerances. It is thus necessary to determine manufacturing tolerances to verify, during the process execution, the generated defects. These defects have to remain in the expected range in order to be certain that the produced part is suitable.

These tolerances must respect two requirements:
- Only the active surfaces in a set-up will be toleranced in this set-up. Active surfaces are positioned and manufactured surfaces in the set-up.
- For ISO tolerances, positioned surfaces cannot be toleranced but only be a part of the datum system.

The tolerance analysis has been performed by the way of constrained optimisation leading to the determination of the worst cases. The result of this analysis is the starting point of the tolerance synthesis:
- First, the study of the influence of each defect parameters on the analysis result allows determining the key surfaces and defects categories (orientation or position) for each surface.
- Second, the set of defects lead to a proposition of a set of ISO manufacturing tolerances suitable to limit their variation.
- The last step consists in the verification of the set of ISO manufacturing tolerances proposed and the determination of the suitable values.

2. LITERATURE REVIEW

In a close field of our work, [Zhou et al, 2003] use a vectorial model to describe defects relative to a nominal model. They gather the modeled defects in a state vector named x(k) for the set-up k. They propose that each manufacturing set-up generates defects which are classified in three categories:
- Positioning defects that correspond to the positioning of a perfect workpiece in an imperfect part-holder,
- Machining defects that correspond to the defects of the machined surfaces relative to the machine-tool,
- Reference defects corresponding to the positioning defects of an imperfect part in a perfect part-holder.

They then use matrices operators to combine the defects and to determine the state vector at the end of the set-up x(k+1).

[Benea et al, 2001] propose a graph based model of the manufacturing process in which the positioning surfaces are underlined with arrows and the machined surfaces by letters "ij". "ij" means that the designated surface has been machined in step j of the set-up i. In the manufacturing process, the positioning defects ($\varepsilon$), the machining defects ($\delta$) and the reference frame of the machine-tool of each set-up are modelised. Dotted lines connect the part surfaces which are constrained by functional tolerances. Those graphs allow, for each analysed tolerances, to determine the influential surfaces and to propose manufacturing tolerances. Then, the analysis of the manufacturing tolerances is led using a small displacement torsors (SDT) model.

[Laperrière et al., 2000] develop a tolerance chart method using successive pairs of surfaces. They classify these pairs of surfaces in internal pairs (belonging to the same part) or kinematic pairs (linking two parts). They write the position of a surface relative to the other one under matrix form. So, the chain of surfaces is described by multiplying transformation matrix describing the relative position of two surfaces.

$$T_0^{6n} = T_0^6 \cdot T_6^{12} \ldots T_{6n-6}^{6n}$$

This equation becomes, by derivation, the deviation matrix of the surfaces relative to the nominal one. Using this model, it is then possible to analyse tolerances, or, by inversion of the matrix, to do the synthesis of tolerances.

[Thiebaut, 2001] proposes an analysis and synthesis method based on an assembly model in which the defects are described relative to the nominal model of the part and the nominal model of the mechanism. These defects are expressed by means of a SDT. The functional requirements are expressed by a SDT the parameters of which are limited. Then, Thiebaut analyses all the possible assembly configurations and check if, in the worst case, the requirement is reach. He proposes a not finalized method of tolerances synthesis. From the model of the mechanism and the knowledge of the influential surfaces obtained by the analysis, the proposed method allows driving the designer in the choice of the functional tolerances.

[Anselmetti, 2000] and [Anselmetti et al, 2003] propose a method of transfer named CLIC. This method is based on the concepts of the standard ISO. It allows the determination of the ISO tolerances and the values of them (tolerance synthesis) on each part to guarantee the assemblability and the respect for the functional requirements. The studied mechanism is represented in the form of positioning tables, this according to the assembly sequence. A set of rules allows deducing the ISO tolerancing of each part of the mechanism. Afterward, a transfer method is proposed for manufacturing tolerances named TZT (Transfer of Zone of Tolerance). This method uses the positioning tables, a table of 3D transfer of tolerances and six rules to make the qualitative synthesis of manufacturing tolerances.

In this paper, our ambition is to propose a more generic approach than those described above, including both the description of the process plan, the defects generated by the manufacturing process, and the functional tolerances analysed.

## 3. THE MMP MODEL

This paragraph presents the concept of Model of Manufactured Part (MMP) we developed in previous papers [Villeneuve et al, 2004][Vignat et al, 2006].

The geometrical model we use to describe the MMP is based on the definition proposed in [Bourdet et al, 1995] and [Thiebaut, 2001] for the assembly analysis and extended by [Villeneuve et al, 2001] and [Vignat et al, 2003] for the manufacturing process analysis. This model is based on an ideal part that is a part made up of perfect form surfaces deviated relative to their perfect (nominal) position. The geometrical description of the MMP is based on the nominal model of the part than can be issued from a CAD model. This nominal model is composed of:
- a global coordinate system which can be built on associated surfaces
- a set of nominal surfaces with, for each surface:
    - the type of surface

- a local coordinate system
- the boundaries (edges and vertices)

The associated part is described relative to the nominal one by the deviation of each surface. The real surfaces are associated to ideal surfaces using a usual criterion like least square. In the MMP, the deviation of these associated surfaces is described relative to the nominal one. This deviation is expressed by a small displacement torsor (SDT) whose structure depends on the surface type. For example, for the plane 6 of the part figure 1, in a local coordinate system (with Z local axis normal to the plane), the torsor is described figure 1. The 3 non-null values represent the potential defects of the surface and the 3 null values represent the invariant degrees of the surface.

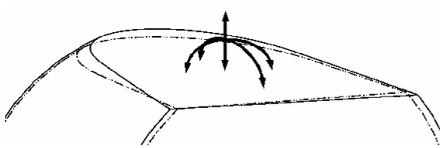

*Figure 1; Plane 6 displacement*

This model is also used to describe the process, that is, for each set-up, the associated part-holder, the part positioning and the machining operations.

The MMP (see figure 3) is a representation of the set of produced parts including description of the process in term of geometrical deviations. It particularly describes the defects generated by the process, classifies these defects and indicates the limits on these one. To illustrate the concepts presented in this paper, an example of a realised part represented figure 2 will be used and a functional tolerance will be analysed.

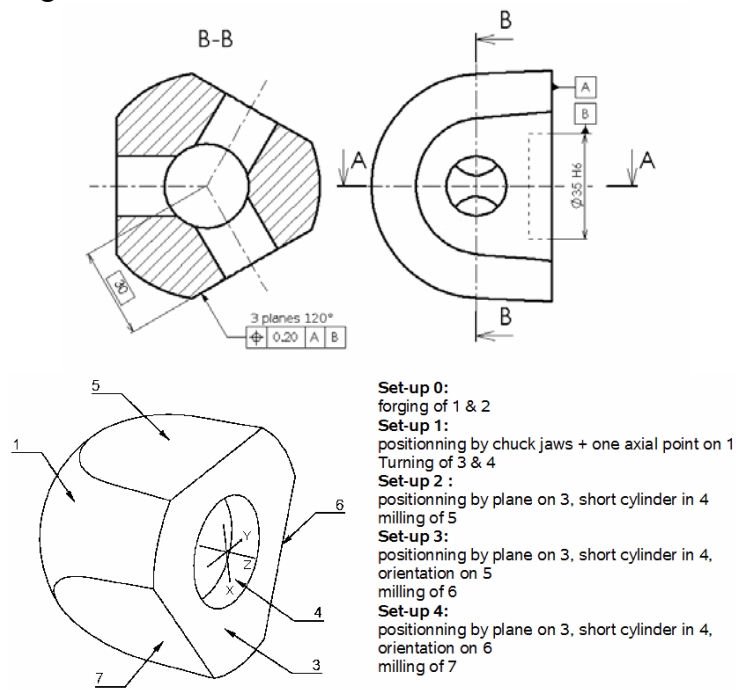

*Figure 2; Example of part and process plan*

The defects generated by a machining process are considered to be the result of two independent phenomenons: the positioning one and the machining one, accumulated over the successive set-ups. The combination of these two phenomenons can be made by summation; the result is the deviation of a realized surface relative to the nominal part.

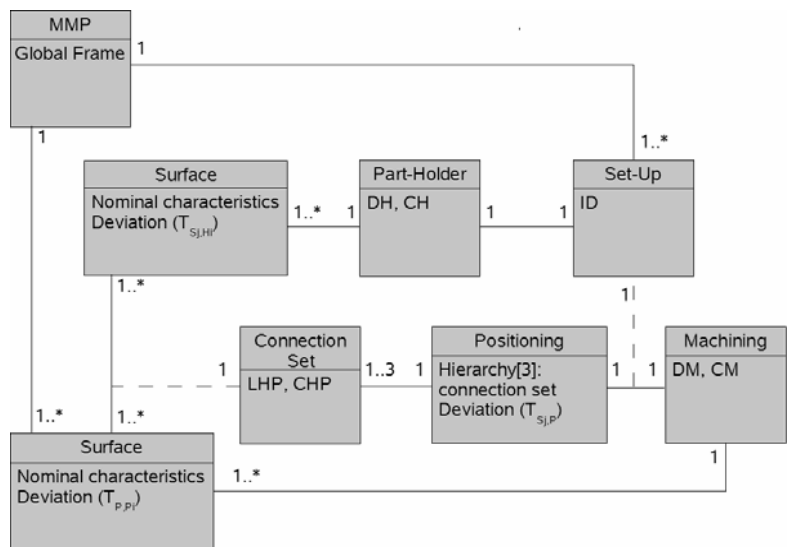

*Figure 3; Structure of the MMP*

The positioning deviation is the deviation of the nominal part relative to the nominal machine. The part positioning operation on the part-holder is realized by a set of hierarchically organized elementary connections [Villeneuve et al, 2005]. Each set determine a part of the 6 degrees of freedom of the MMP in the 3D space. The positioning deviation is expressed by a small displacement torsor $T_{Sj,P}$ for set-up j. This deviation is function of the MMP surfaces deviation generated by the previous set-ups, the part-holder surfaces deviations and the link part-holder/part surfaces.

The part-holder surfaces deviations are described by small displacement torsors $T_{Sj,Hk}$ for surface k of the part-holder in set-up j . The parameters of these torsors (DH) are saved in the part-holder object of the MMP and limited by constraints (CH) representative of the part-holder quality. These constraints are limiting one or a set of parameters together.

An elementary connection links an associated surface of the part-holder and a surface of the MMP. The relative position of these 2 surfaces is described by a small displacement torsor $T_{Hk,Pi}$ called link torsor. The 6 parameters of this torsor are the link parameters (LHP). The LHP values depend on connected surfaces contact condition and can be calculated for each part/part-holder pair using positioning rules expressed by constraints (CHP). Two types of contacts are defined [Dantan, 2000]: the floating and the slipping one. For floating contacts, the only constraint is non-penetration of the part in the part-holder and is determined on the boundary edge and vertex. For slipping contact the part is pushed on the part holder by a clamping strength. The constraint associated to this type of contact is a positioning function to maximise combined to non-penetration conditions. The connection data are saved in the connection sets objects of the set-up.

The deviation of the surface machining operation is described relative to the nominal machine by a small displacement torsor $T_{Sj,Pi}$ for the surface i realised in set-up j. This torsor is includes deviation of the surface swept by the tool and cutting local deformations. The parameters of this torsor represent machining deviations (DM). They are limited by constraints (CM) representing the machines and tools capabilities. These data are stored in the machining object of the MMP.

For each surface of the part, the positioning and machining effects are added and its deviation relative to the nominal part is determined and expressed by a small displacement torsor $T_{P,Pi}$ for surface i of the part as described in figure 4 for surface 6 of the example figure 2. This deviation is stored in the surface object of the MMP.

$$T_{P,P6} = \left\{ \begin{array}{cc} Urz_{3S2} + ry_{5S3} + lry_{5S3} + rx_5 + rx_6 & 0 \\ \begin{array}{c} 0,86\,lrx_{10S1} - 0,86\,lrx_{3S3} + 0,5\,lry_{10S1} + 0,5\,lry_{3S3} \\ -0,86\,rx_1 + 0,86\,rx_{10S1} - 0,86\,rx_3 - 0,86\,rx_{3S3} \\ -0,5\,ry_1 + 0,5\,ry_{10S1} - 0,5\,ry_3 + 0,5\,ry_{3S3} + ry_6 \end{array} & 0 \\ 0 & \begin{array}{c} -1,73\,lrx_{10S1} - 6,49\,lrx_{3S3} - lry_{10S1} + 3,75\,lry_{3S3} \\ +0,5\,ltx_{10S1} + 0,5\,ltx_{4S3} - 0,86\,lty_{10S1} - 0,86\,lty_{4S3} \\ -8,66\,rx_1 - 1,73\,rx_{10S1} - 6,49\,rx_3 - 6,49\,rx_{3S3} \\ +2,16\,rx_4 - 5\,ry_1 - ry_{10S1} - 3,75\,ry_3 \\ +3,75\,ry_{3S3} + 1,25\,ry_4 - 0,5\,tx_1 + 0,5\,tx_{10S1} \\ -0,5\,tx_4 + 0,5\,tx_{4S3} + 0,86\,ty_1 - 0,86\,ty_{10S1} \\ +0,86\,ty_4 - 0,86\,ty_{4S3} + tz_6 \end{array} \end{array} \right\}$$

*Figure 4; Deviation of a surface of the MMP*

In [Vignat et al, 2005] it is proposed to model the functional tolerance to be analysed by a virtual gauge (see figure 5). The virtual gauge is a perfect part made up of positioning surfaces and tolerance zones surfaces.

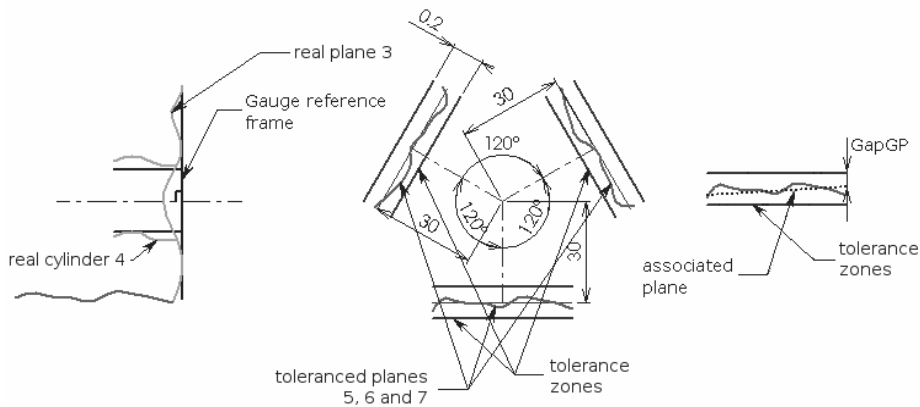

*Figure 5; Virtual gauge for analyzed tolerance*

This virtual gauge is assembled with the MMP and the verification is made by the way of a signed distance GapGP (see figure 5 on the right) measuring the position of the MMP surface relative to the tolerance zone bounding surfaces. The worst case searching is made by the mean of the optimisation function (1).

$$\underbrace{\underset{DM,DH,LHP}{\overset{CM,CH,CHP}{\text{Min}}}}_{(1.a)} \underbrace{(\underset{LGP}{\overset{CGP}{\text{Max}}}(\underset{j}{\text{Min}}(GapGP_j)))}_{(1.b)} \qquad (1)$$

(1.b) is the mathematical expression of the tolerance to be verified therefore it represent the signed distance GapGP measured between the gauge tolerance zone and the MMP. This measure is made while the link parameters LGP conform to assembly constraints CGP defined in accordance with the tolerancing standard. (1.a) is the expression of the worst case search, therefore the worst combination of manufacturing defects DM, DH, LHP within the estimated (or measured) range of variation expressed by the constraints CM, CH, CHP.

## 4. TOLERANCE SYNTHESIS

The tolerance analysis step allows concluding about process feasibility for the part to be realised. It gives also information about each defect parameters influence by the way of influence coefficient presented table 1. These coefficients are in fact the gradient of the parameters in the optimisation problem.

| Set-up 1 | | | Set-up 2 | | | Set-up 3 | | | Set-up 4 | | |
|---|---|---|---|---|---|---|---|---|---|---|---|
| Machining | | | Positioning | | | Positioning | | | Positioning | | |
| Plane 3 | $rx_3$ | | Plane 3S2 | $rx_{3S2}$ | | Plane 3S3 | $rx_{3S3}$ | 34,64 | Plane 3S4 | $rx_{3S4}$ | 34,64 |
| | $ry_3$ | | | $ry_{3S2}$ | 40 | | $ry_{3S3}$ | 20 | | $ry_{3S4}$ | 20 |
| | $rz_3$ | | | $tz_{3S2}$ | | | $tz_{3S3}$ | | | $tz_{3S4}$ | |
| Cylinder 4 | $ra_4$ | | Cylinder 4S2 | $ra_{4S2}$ | | Cylinder 4S3 | $ra_{4S3}$ | | Cylinder 4S4 | $ra_{4S4}$ | |
| | $rx_4$ | | | $rx_{4S2}$ | | | $rx_{4S3}$ | | | $rx_{4S4}$ | |
| | $ry_4$ | | | $ry_{4S2}$ | | | $ry_{4S3}$ | | | $ry_{4S4}$ | |
| | $tx_4$ | | | $tx_{4S2}$ | 1 | | $tx_{4S3}$ | 0,5 | | $tx_{4S4}$ | 0,5 |
| | $ty_4$ | | | $ty_{4S2}$ | | | $ty_{4S3}$ | 0,87 | | $ty_{4S4}$ | 0,86 |
| | | | Machining | | | Ry5S3 | $ry_{5S3}$ | 25 | Ry5S4 | $ry_{6S4}$ | 25 |
| | | | Plane 5 | $rx_5$ | 25 | Machining | | | Machining | | |
| | | | | $ry_5$ | 25 | Plane 6 | $rx_6$ | 25 | Plane 7 | $rx_7$ | 25 |
| | | | | $tz_5$ | 1 | | $ry_6$ | 25 | | $ry_7$ | 25 |
| | | | | | | | $tz_6$ | 1 | | $tz_7$ | 1 |

*Table 1; defect parameters influence on the analysed tolerance*

The influential parameters names indicate the concerned surfaces and type of defects for each surface (orientation or position). It is possible to determine:
- The concerned surface by the number i in the i or iSj index
- How the surface is concerned in the set-up, index iSj indicate that it is a positioning surface while index i indicate that it is a machined surface

- The type of defect contributing to the set-up: rx, ry and rz name indicate orientation defects and tx, ty and tz position defects.

From table 1 it is possible to determine for the studied example that, in set-up 3, positioning surface plane 3 is concerned for orientation, cylinder 4 for orientation around global x and y axis (see figure 2) and plane 5 for orientation around global z axis. Concerning machined surfaces, plane 6 is concerned for orientation and position.

For each set-up, the active surfaces and related defects have been identified by the mean of influential parameters. It is then possible to propose ISO specifications respecting the principles stated at the beginning of this chapter:
- Only the active surfaces in a set-up will be toleranced in this set-up. Active surfaces are positioned and manufactured surfaces in the set-up.
- For ISO tolerances, positioned surfaces cannot be toleranced but only be a part of the datum system.

In this step, a datum system and one or more specifications to limit the identified defects have to be proposed. If parameters identify positioning surfaces, these surfaces are chosen to compose the reference frame. The hierarchical order of the reference frame is determined according to the hierarchical order of positioning surfaces in the set-up. Surfaces identified as machined in the set-up are specified in orientation, localization or both according to the defect parameters determined. Translation and rotation parameters can be limited together by location specification while rotation parameters only can be limited by orientation specification.

For the present example, the proposition for set-up 3 is represented figure 6. First a reference frame on 3, 4 and 5 is defined because the parameters identify the influence of these 3 surfaces and the hierarchy in the positioning set is 3, 4 and 5. The parameters identify only one machined surface (plane 6) and translation and rotation defects. It is thus necessary to specify the surface 6 by specification of localization.

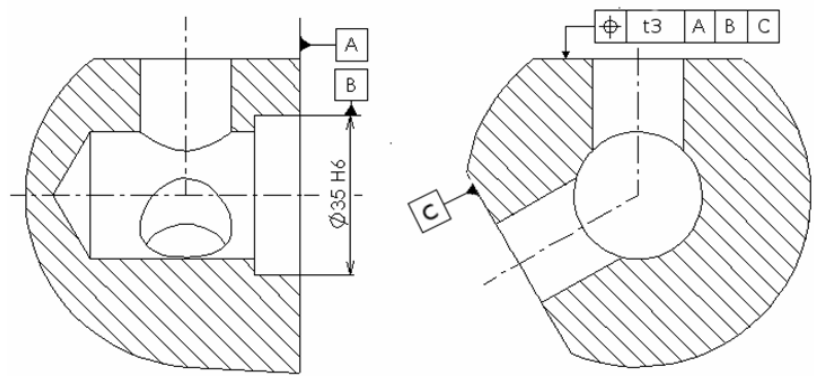

*Figure 6; proposition of a set of manufacturing specifications for set-up 3*

For the determination of the tolerances values and the checking of the relevance of the proposed set of manufacturing specifications, the solution consists in using a numerical algorithm of optimization identical to that of the analysis stage.

This algorithm will be used to solve the optimisation problem expressed by (2).

$$\underset{\substack{DM,DH,LHP,LMGP \\ \underbrace{\phantom{DM,DH,LHP,LMGP}}_{(2.a)}}}{\overset{CMGP,GapMGP>0}{Min}} (\underbrace{\overset{CGP}{\underset{LGP}{Max}}(\underset{j}{Min}(GapGP_j)))}_{(2.b)} \qquad (2)$$

Like (1.b), (2.b) is the mathematical expression of the functional tolerance to be verified therefore it represent the signed distance GapGP measured between the gauge tolerance zone and the MMP (see figure 3 on the right). If the value of (2.b) remains positive, then the studied functional specification remain checked for the worst combination of the manufacturing defects parameters DM, DH and LHP under constraint of respect of the manufacturing specifications expressed by GapMGP>0 (see (2.a)). Like functional specification, manufacturing ones are modelled by virtual gauge which are assembled with the model of workpiece (intermediate MMP at the end of the current set-up). The compliance with the manufacturing tolerance is measured by a signed distance GapMGP. GapMGP>0 imply that there is not penetration between the bounding surfaces of the tolerance zone (TZ) of the virtual gauge representing the manufacturing specification and the toleranced surface and thus the compliance with this specification. The link parameters LMGP between the MWP and the virtual manufacturing gauge are constrained by assembly conditions CMGP defined according the tolerancing standard.

By this method it is possible to determine:
- if the proposed set of manufacturing tolerances is complete. If not, the optimisation problem will diverge to negative values,
- the value of manufacturing tolerances that guarantees the compliance with the functional specification. In this case, the optimisation problem reach a positive value,
- the unnecessary manufacturing specifications. The elimination of the constraints resulting form these tolerances will not modify the result of the optimisation problem.

5. CONCLUSION

The tolerance analysis by the way of an optimisation method allows determining, by a numerical mean, key surfaces and defects.

The analysis of the key parameters allows proposing a set of manufacturing specifications limiting the concerned defects.

An optimisation method is then used to verify if the proposed set of specifications is complete to limit the defects and does not over constrain them. This optimisation step is also necessary to verify the chosen tolerances values associated with the specifications.

This method implies the use of a trial error algorithm that can need many trials steps. On the other hand, this is a numerical method allowing its deployment in CAD/CAM software.